\documentstyle[12pt]{article}
\newcommand{\bea}{\begin{eqnarray}}
\newcommand{\be}{\begin{equation}}
\newcommand{\eea}{\end{eqnarray}}
\newcommand{\ee}{\end{equation}}
\def\nn{\nonumber}

\let\ssection=\section
\renewcommand{\section}{\setcounter{equation}{0}\ssection}

\def\a{\alpha}
\def\b{\beta}

\def\d{\delta}
\def\e{\epsilon}

\def\g{\raisebox{.4ex}{$\gamma$}}

\def\k{\kappa}
\def\l{\lambda}

\def\p{\pi}
\def\q{\theta}

\def\t{\theta}

\def\r{\rho}

\def\t{\tau}

\def\G{\Gamma}
\def\J{\Psi}
\def\L{\Lambda}

\def\ca{{\cal A}}

\def\cd{{\cal D}}

\def\cg{{\cal G}}
\def\ch{{\cal H}}

\def\cl{{\cal L}}
\def\cm{{\cal M}}

\def\car{{\cal R}}

\def\cv{{\cal V}}

\def\hg{{\tilde{g}}}
\def\lk{{\scriptscriptstyle (k)}}
\def\lkmi{{\scriptscriptstyle (-k)}}
\def\lel{{\scriptscriptstyle (l)}}
\def\lm{{\scriptscriptstyle (m)}}
\def\l0{{\scriptscriptstyle (0)}}
\def\len{{\scriptscriptstyle (n)}}
\def\lkl{{\scriptscriptstyle (k+l)}}

\def\lkm{{\scriptscriptstyle (k+m)}}

\def\lkmn{{\scriptscriptstyle (k+m+n)}}

\begin{document}
\pagestyle{empty} \large \noindent . \vspace*{5mm} \\
CGPG-94/4-6\\April 1994\\
\begin{center} \LARGE \bf \vspace*{35mm}
Real formulations of complex gravity and a complex formulation of real
gravity\\
\vspace*{20mm} \large \bf
Peter Peld\'{a}n
\footnote{Email address: peldan@phys.psu.edu}\\
\vspace*{5mm} \large
Center for Gravitational Physics and Geometry\\
The Pennsylvania State University, University Park, PA 16802, USA\\
\vspace*{25mm} \Large \bf
Abstract\\
\end{center} \normalsize
Two gauge and diffeomorphism invariant theories on the Yang-Mills phase space
are studied. They are based on the Lie-algebras $so(1,3)$ and
$\widetilde{so(3)}$ -- the loop-algebra of $so(3)$. Although the theories are
manifestly real, they can both be reformulated to show that they describe
 complex gravity and an infinite number of copies of complex
gravity, respectively. The connection to real gravity is given. For these
theories, the reality conditions in the conventional Ashtekar formulation are
represented by normal constraint-like terms.
\\ PACS number: 04.20.Fy

\newpage \pagestyle{plain}
\section{Introduction}
In the search for a generalization of the Ashtekar variables \cite{Ash1}
to other gauge
groups, it has previously been shown \cite{pp2+1},
\cite{arbgg}, \cite{subenoy} that there
exist an arbitrary gauge group generalization in both (2+1) as well as
(3+1)-dimensions. Moreover, for both these cases, the generalized theories
were shown to produce the conventional Yang-Mills theory when expanded around
de Sitter spacetime. However, the problem with the (3+1)-dimensional model is
that the theory is required to be complex in order to include Lorentzian
spacetimes, and no reality conditions have yet been found. Another
expectation one may have on a unified description of gravity and Yang-Mills
theory, is that the theory itself should be somewhat restrictive regarding
what gauge groups that could be used. This is not the case for the models
presented in \cite{pp2+1} and \cite{subenoy}, these constructions allow all
gauge groups whose Lie-algebras admit a non-degenerate invariant bilinear
form. Thus, motivated by the desire to solve these two problems, I have, in
this paper, studied another type of generalization of the Ashtekar
formulation. The method used in \cite{subenoy} was to replace the structure
constant of $so(3)$ by the three-dimensional internal Levi-Civita symbol,
which then was generalized to be a well defined object for higher dimensional
Lie-algebras. In this paper, I instead keep the structure constants in the
theory, and use only Lie-algebras whose structure constants satisfy an
identity of the type
\be f_{IJK}f^K{}_{LM}\sim \sum _{i,j}g^i_{I[L}g^j_{M]J} \ee
where the $g^i_{IJ}$ are invariant bilinear forms of the Lie-algebra. I show
that it is possible to find a gauge and diffeomorphism invariant theory based
on two such Lie-algebras: $so(1,3)$ and the loop-algebra $\widetilde{so(3)}$.
However, as can be shown by reformulating the theories, both these models
describe nothing more than complex gravity. The $so(1,3)$ theory is exactly
equivalent to complex gravity, while the $\widetilde{so(3)}$ theory
corresponds to an infinite number of copies of complex gravity. The
interesting fact about these formulations is perhaps that they represent a
way of getting real formulations of complex gravity, and they may possibly be
used to shed some light over the reality condition-problem in Ashtekar's
variables.\\ \\
The models presented here are constructed in (3+1)-dimensions for
the Lie-algebras $so(1,3)$ and $\widetilde{so(3)}$. But it is straightforward
to generalize the formulation to other dimensions $\geq (2+1)$ and for the
Lie-algebras $so(4)$, $so(2,2)$, $so(1,2)$ -- and isomorphic ones --
as well as their loop-algebras.

\section{What is the problem with arbitrary gauge groups?}\label{problem}
The problem of using arbitrary gauge groups for the Ashtekar Hamiltonian
comes from the constraint algebra; the algebra fails to close in general.
In the Ashtekar formulation, the first class constraints are
\bea \cg ^I&:=&\cd _aE^{aI}=\partial _aE^{aI}+f^I{}_{JK}A_a^JE^{aK}\approx 0
\label{Gauss} \\
\ch _a&:=&\frac{1}{2}\e _{abc}E^{bI}B^c_I\approx 0 \label{vector} \\
\ch &:=& \frac{1}{4}\e _{abc}f_{IJK}E^{aI}E^{bJ}B^{cK} \approx 0
\label{Ham} \eea
They are called, the Gauss law, the vector constraint, and the Hamiltonian
constraint, respectively. The conventions are: $a, b, c, ...$ are spatial
indices on the hypersurface, $I,J,K, ...$ are Lie-algebra indices in the
vector representation. The structure constant is denoted $f_{IJK}$, and
Lie-algebra indices are raised and lowered by an invariant bilinear form on
the Lie-algebra. (For the conventional Ashtekar Hamiltonian, which together
with a set of reality conditions describes real Einstein gravity, the
Lie-algebra
is $so(3;C)$ and the bilinear form is $\d _{IJ}$.) The basic conjugate fields
are a gauge connection $A_{aI}$ and "the electric field" $E^{aI}$. They
satisfy: $\{A_{aI}(x),E^{bJ}(y)\}=\d ^J_I\d ^b_a \d ^3(x-y)$. $B^{aI}$ is the
"magnetic field": $B^{aI}:=\e ^{abc}F_{bc}^I=\e ^{abc}(2 \partial _bA_c^I +
f^I{}_{JK}A_b^JA_c^K)$. The totally antisymmetric Levi-Civita tensor
densities are $\e ^{abc}$ and $\e _{abc}$. They are taken to have $\e
^{123}=\e _{123}=1$ in every coordinate system.

Since $\cg ^I$ and $\tilde{\ch}_a:=\ch _a -A_{aI}\cg ^I$ easily can be shown
to generate gauge transformations and spatial diffeomorphisms, there can be
no problems with the Poisson brackets containing these constraints, as long
as all constraints are gauge- and diffeomorphism covariant. The only
potential obstruction to get a closed constraint algebra, thus comes from
$\{\ch [N],\ch [M]\}$. A straightforward calculation gives
\be \{\ch [N],\ch [M]\}= \frac{1}{2}\int _{\Sigma}d^3x
\left ((N\partial _aM - M\partial
_aN)E^{aI}E^{bJ}\e _{bcd}E^{cK}B^{dL}f_{IJM}f^M{}_{KL}\right )
\label{HHgen}\ee

A right-hand side which in general is {\it not} a linear combination of
constraints. For $SO(3)$ or locally isomorphic groups,
the structure constant equals the epsilon symbol, and therefore the
right-hand side above reduces to the vector constraint smeared over some
test function. For arbitrary gauge groups, no such simplification occurs, and
the constraint algebra therefore fails to close. Thus, it seems that the
requirement on the Lie-algebra is that its structure constants must
satisfy $f_{IJ}{}^Kf_{KLM}\sim g_{IL}g_{JM}-g_{IM}g_{JL}$ in order to make
the constraints in eq. (\ref{Gauss})-(\ref{Ham}) form a first class set.
Here, $g_{IJ}$ is an invariant bilinear form on the Lie-algebra.
This is a severe restriction on the Lie-algebra, and in fact I only know
of two algebras that satisfy this: $so(3)$ and $so(1,2)$.
However, in this paper, I will show how one can use a slightly
weaker condition: suppose we have an Lie-algebra that satisfies
\be f_{IJK}f^K{}_{LM}\sim \sum _{i,j}g^i_{I[L}g^j_{M]J} \label{genid}\ee
where $g^i_{IJ}$ are invariant bilinear forms of the Lie-algebra. The indices
$i$ and $j$ just label the different bilinear forms. Using this in
(\ref{HHgen})
above, we see that the right hand side becomes a linear combination of the
vector constraint and "vector constraint-like" terms. By introducing the
"vector constraint-like" terms as new constraints in the theory, and continue
to check the algebra, one soon notices that the new constraints also produce
new "Hamiltonian constraint-like" terms, which also have to be included in
the set of constraints. However, if the Lie-algebra is such that the product
of the bilinear forms also is invariant, the procedure stops here, and we
have a first class set. That is, I require
\be g^i_{IJ}g^{jJ}{}_K=\sum _k a_k g^k_{IK} \label{g2} \ee
where I have selected one of the bilinear forms, say $g^0_{IJ}$, and its
inverse to raise and lower Lie-algebra indices.

I do not
know of any general theorems concerning what kind of Lie-algebras that
satisfy these requirements. In the following sections, two different examples
of this construction will be presented. They are based on $so(1,3)$ and the
loop-algebra $\widetilde{so(3)}$.

\section{The $SO(1,3)$ theory}\label{so(1,3)}
In this section, I will present an Ashtekar-like theory based on the
Lie-algebra $so(1,3)$. The construction also works for the algebras $so(4)$
and $so(2,2)$. First I describe the Lie-algebra in subsection
\ref{algso(1,3)}.
In subsection \ref{Hso(1,3)}, the Hamiltonian is presented and a constraint
analysis is performed. In subsection \ref{split}, the theory is split up into
self dual and antiself dual parts, and I show that the real $so(1,3)$ theory
is equivalent to the complex $so(3)$ Ashtekar formulation. Finally, in
subsection \ref{Lso(1,3)}, two manifestly covariant and real Lagrangians for
the $so(1,3)$ theory are derived.
\subsection{The $so(1,3)$ algebra} \label{algso(1,3)}
Before introducing the theory, I will give a short description of the
Lie-algebra. First, a basis is chosen for the fundamental representation:
\be \begin{array}{lll} T_{1A}{}^B=\left (\begin{array}{cccc}0 & 1 & 0 & 0 \\
1 & 0 & 0 & 0 \\ 0 & 0 & 0 & 0 \\ 0 & 0 & 0 & 0 \end{array}\right ) &
T_{2A}{}^B=\left (\begin{array}{cccc}0 & 0 & 1 & 0 \\
0 & 0 & 0 & 0 \\ 1 & 0 & 0 & 0 \\ 0 & 0 & 0 & 0 \end{array}\right ) &
T_{3A}{}^B=\left (\begin{array}{cccc}0 & 0 & 0 & 1 \\
0 & 0 & 0 & 0 \\ 0 & 0 & 0 & 0 \\ 1 & 0 & 0 & 0 \end{array}\right ) \\
T_{4A}{}^B=\left (\begin{array}{cccc}0 & 0 & 0 & 0 \\
0 & 0 & 1 & 0 \\ 0 & -1 & 0 & 0 \\ 0 & 0 & 0 & 0 \end{array}\right )
& T_{5A}{}^B=\left (\begin{array}{cccc}0 & 0 & 0 & 0 \\
0 & 0 & 0 & -1 \\ 0 & 0 & 0 & 0 \\ 0 & 1 & 0 & 0 \end{array}\right )&
T_{6A}{}^B=\left (\begin{array}{cccc}0 & 0 & 0 & 0 \\
0 & 0 & 0 & 0 \\ 0 & 0 & 0 & 1 \\ 0 & 0 & -1 & 0 \end{array}\right )
\end{array} \label{basis} \ee
where $A, B, ...$ are four-dimensional Lorentz-indices. I will also
use $I, J, K, ...$ as six-dimensional indices in the vector
representation. $T_1, T_2$ and $T_3$ are the boost-generators, and $T_4, T_5,
T_6$ are the generators of rotations.
Using this representation, it is straightforward to calculate the commutator
algebra of
these basis elements:
\be [T_I,T_J]=f_{IJ}{}^KT_K \ee
where
\bea
&&f_{12}{}^4=f_{31}{}^5=f_{23}{}^6=f_{45}{}^6=f_{64}{}^5=f_{56}{}^4=f_{14}{}^2
=f_{51}{}^
3=f_{42}{}^1=f_{26}{}^3=f_{35}{}^1=f_{63}{}^2=1 \nn \\
&&f_{16}{}^i=f_{25}{}^i=f_{34}{}^i=0  \eea
Now, I need to introduce a non-degenerate invariant bilinear form -- or a
scalar product
-- in
the Lie-algebra. Invariance here means, invariance under conjugation (gauge
transformations). If the bilinear form is denoted $<A,B>$ for $A$ and
$B$ belonging to the Lie-algebra, the invariance requirement becomes:
\be <[C,A],B>+<A,[C,B]>=0 \ee
for all elements $A$, $B$ and $C$ belonging to the Lie-algebra.
Or, expressing $A$, $B$ and $C$ in the basis $T_I$ above: $A=A^IT_I$ etc.,
the invariance condition reads
\be f_{IJ}{}^Kg_{KL}+f_{IL}{}^Kg_{KJ}=0 \ee
where I have defined $g_{IJ}:=<T_I,T_J>$. With the $f_{IJ}{}^K$'s given
above, it is easy to solve this equation for $g_{IJ}$, in this
representation.
 The solution is that there
exist only two different non-degenerate invariant bilinear forms
\be \begin{array}{ll} g_{IJ}=\left ( \begin{array}{cccccc} -1 & 0 & 0 & 0 & 0
& 0 \\ 0 & -1 & 0 & 0 & 0
& 0 \\ 0 & 0 & -1 & 0 & 0
& 0 \\ 0 & 0 & 0 & 1 & 0
& 0 \\ 0 & 0 & 0 & 0 & 1
& 0 \\ 0 & 0 & 0 & 0 & 0
& 1 \end{array} \right ) &
g^\ast _{IJ}=\left ( \begin{array}{cccccc} 0 & 0 & 0 & 0 & 0
& 1 \\ 0 & 0 & 0 & 0 & 1
& 0 \\ 0 & 0 & 0 & 1 & 0
& 0 \\ 0 & 0 & 1 & 0 & 0
& 0 \\ 0 & 1 & 0 & 0 & 0
& 0 \\ 1 & 0 & 0 & 0 & 0
& 0 \end{array} \right ) \end{array} \label{forms}\ee
Of course, any linear combination of these matrices will also be an invariant
bilinear form.
The reason why this Lie-algebra allows two different "group-metrics" can be
traced back to the fact that it splits into two subalgebras $so(1,3;C)=
so(3,C)\oplus so(3,C)$. If one wants an explicit definition of
 $g_{IJ}$ it is simply given by the Cartan-Killing form
$-\frac{1}{4}f_{IK}{}^Lf_{JL}{}^K$ or the matrix-trace
$-\frac{1}{2}Tr(T_IT_J)$. The explicit definition of $g^\ast _{IJ}$ can be
found by first defining the dual basis $T^\ast _{IA}{}^B:=\frac{1}{2}\e
_A{}^{BCD}T_{ICD}$ and then take the matrix-trace
$-\frac{1}{2}Tr(T_IT_J^\ast)$. Here, I have introduced the totally
antisymmetric $\e ^{ABCD}$ and the Lorentz-indices are raised and lowered
with the
Minkowski metric $\eta _{AB}=diag(-1,1,1,1)$. Thus, in the fundamental
representation, $g_{IJ}$ corresponds to "tracing" with two Minkowski metrics,
while $g^\ast _{IJ}$ represents the $\e _{ABCD}$-trace.

Now, what is interesting about this Lie-algebra, for our purposes, is that
its structure constants satisfy the identity
\be f_{IJ}{}^Kf_{LK}{}^M=\d ^M_{[I}g_{J]L} + g^\ast _{[J}{}^Mg^\ast _{I]L}
\label{id}\ee
where I have defined $g^\ast _I{}^J:=g^{JK}g^\ast _{IK}$ and $g^{IJ}$ is the
inverse of $g_{IJ}$. From now on I will raise and lower Lie-algebra indices
 with $g_{IJ}$ and $g^{IJ}$. If an index is lowered with $g^\ast
_{IJ}$, the new object will be denoted with an extra $\ast $: {\it e.g.}
$f^\ast
_{IJK}:=f_{IJ}{}^Lg^\ast _{KL}$. Note that the identity (\ref{id}) is
exactly of the type (\ref{genid}), and that the product of two bilinear
forms always will be an invariant bilinear form.
Now, I summarize all the important
definitions and relations for this algebra:
\be \begin{array}{ll} g_{IJ}:=-\frac{1}{2}Tr(T_IT_J)=-\frac{1}{2}T_{IA}{}^B
T_{JB}{}^A & g^\ast _{IJ}:=-\frac{1}{2}Tr(T_IT^\ast _J)=-\frac{1}{4}\e
^{ABCD} T_{ICD}T_{JBA} \\ & \\ g^{IJ}g_{JK}=\d ^I_K &
g^{\ast IJ}:=g^{IK}g^{JL}g^\ast _{KL}\\ & \\  g^{\ast IJ}g^\ast
_{JK}=-\d ^I_K & [T_I,T_J]=f_{IJ}{}^KT_K \\ & \\
f_{IJL}:=f_{IJ}{}^Kg_{KL}=-Tr(T_IT_JT_L) & f^\ast_{IJK}:=f_{IJ}{}^Lg^\ast
_{KL} \\ & \\ f^{\ast \ast}_{IJK}=-f_{IJK}  &
f_{IJ}{}^Kf_{LK}{}^M=\d ^M_{[I}g_{J]L} +
g^\ast _{[J}{}^Mg^\ast _{I]L}  \end{array} \ee
\subsection{The $so(1,3)$ Hamiltonian}\label{Hso(1,3)}
Then, it is time to start building a gauge and diffeomorphism invariant
theory based on this Lie-algebra. The fundamental
phase space variables are chosen to be a connection $A_{aI}$ and its
conjugate momenta $E^{aI}$:
\be
\{ A_{aI}(x),E^{bJ}(y)\}=\d _a^b\d _I^J\d ^3(x-y) \label{FPB} \ee.
More generally, one
could have chosen the right-hand side to be $\d ^b_a(\d ^J_I + a g^{\ast J}
_I)$. However, this Poisson bracket may be brought to the standard one
(\ref{FPB}) by a simple field-redefinition:
$\tilde{E}^{aI}:=\frac{1}{1+a^2}(\d ^I_J - a
g^{\ast I}_J)E^{aJ}$. Now, according to the outline of the constraint algebra
calculation, given in section \ref{problem}, we know that the full set of
constraints we need includes the conventional Ashtekar constraints as well as
the new constraints that are produced by replacing $\d ^I_J$ by $g^{\ast
I}_J$ in the vector and the Hamiltonian constraint. Thus, we have
\bea \cg _I&:=&\cd _aE^a_I=\partial _aE^a_I + f_{IJK}A_a^JE^{aK} \approx 0
\label{G}\\
\ch _a&:=&\frac{1}{2}\e _{abc}E^{bI}B^c_I\approx 0 \label{Ha} \\
 \ch _a^\ast&:=&\frac{1}{2}\e _{abc}E^{bI}B^{\ast c}_I\approx 0 \label{Has} \\
\ch &:=&\frac{1}{4}\e _{abc}f_{IJK}E^{aI}E^{bJ}B^{cK} + \frac{\L}{6} \e
_{abc}f^\ast _{IJK}E^{aI}E^{bJ}E^{cK}\approx 0 \label{H} \\
\ch ^\ast &:=&\frac{1}{4}\e _{abc}f^\ast _{IJK}E^{aI}E^{bJ}B^{cK} -
\frac{\L}{6} \e
_{abc}f^\ast _{IJK}E^{aI}E^{bJ}E^{cK}\approx 0 \label{Hs}
 \eea
Here, I have introduced a cosmological constant term as well, and the
conventions are chosen such that this theory will contain the conventional
Ashtekar formulation for a real $\L$.

The claim is now that these constraints form a first class set, and to prove
that, one simply calculates the constraint algebra. First, to simplify the
calculations, one may note that $\cg ^I$, $\tilde{\ch }_a:=\ch _a -A_{aI}\cg
^I$ and $\tilde{\ch}^\ast _a:=\ch ^\ast _a-A^\ast _{aI}\cg ^I$ generate gauge
transformations, spatial diffeomorphisms and spatial diffeomorphisms times a
duality rotation, respectively:
\bea \d ^{\cg ^I}E ^{aI}&:=&\{E ^{aI},\cg ^J[\L_J]\}=f^I{}_{JK}\L^JE ^{aK}
 \label{a6a}\\
\d ^{\cg ^I}A _a ^I&:=&\{A _a ^I,\cg ^J[\L_J]\}=-\cd _a\L^I  \\
 \d ^{\tilde{\ch}_a}E ^{aI}&:=&\{E ^{aI},\tilde{\ch}_b[N^b]\}\!=\!N^b
\partial _b
 E
^{aI} \!-\! E ^{bI}\partial _b N^a \!+\! E ^{aI} \partial _b N^b\!=\!
\pounds _{N^b}E ^{aI}
 \\
\d ^{\tilde{\ch}_a}A _a ^I&:=&\{A _a ^I,\tilde{\ch}_b[N^b]\}\!=\!N^b
\partial _b
A _a ^I \!+\!
A _{b}^I\partial _a N^b\!=\!\pounds _{N^b}A _a ^I  \label{a9} \\
\d ^{\tilde{\ch}_a^\ast}E ^{aI}&:=&\{E
^{aI},\tilde{\ch}_b^\ast[N^b]\}\!=\!N^b
\partial _b
 E
^{\ast aI} \!-\! E ^{\ast bI}\partial _b N^a \!+\! E ^{\ast aI} \partial _b
N^b\!=\!
\pounds _{N^b}E ^{\ast aI} \label{a10}
 \\
\d ^{\tilde{\ch}_a^\ast }A _a ^I&:=&\{A _a
^I,\tilde{\ch}_b^\ast[N^b]\}\!=\!N^b\partial _b
A _a ^{\ast I} \!+\!
A _{b}^{\ast I}\partial _a N^b\!=\!\pounds _{N^b}A _a ^{\ast I} \label{a11}
\eea
With this knowledge at hand, it is a simple task to calculate all Poisson
brackets containing these constraints. Since $\cg ^I$ is gauge covariant,
, $\ch$ and $\ch ^\ast$
are gauge invariant, and all constraints are diffeomorphism covariant, we get
\bea
\{{\cal G}^I [\Lambda_I],{\cal G}^J[\Gamma_J]\}&=&{\cal G}^K[f_{KIJ}\Lambda^I
\Gamma^J]\label{3.15} \\
\{{\cal G}^I [\Lambda_I],\tilde{{\cal H}}_a[N^a]\}&=&{\cal
G}^I[-\pounds_{N^a}\Lambda^I]\\
\{{\cal G}^I [\Lambda_I],\tilde{{\cal H}}^\ast _a[N^a]\}&=&{\cal
G}^I[-\pounds_{N^a}\Lambda^{\ast I}]\\
\{{\cal G}^I [\Lambda_I],{\cal H}[N]\}&=&0\\
\{{\cal G}^I [\Lambda_I],{\cal H}^\ast[N]\}&=&0\\
\{\tilde{{\cal H}}_a[N^a],\tilde{{\cal H}}_b[M^b]\}&=&\tilde{{\cal H}}_a
[-\pounds_{M^b}
N^a]\\
\{\tilde{{\cal H}}_a[N^a],\tilde{{\cal H}}^\ast _b[M^b]\}&=&\tilde{{\cal
H}}^\ast _a
[-\pounds_{M^b}
N^a]\\
\{\tilde{{\cal H}}^\ast _a[N^a],\tilde{{\cal H}}^\ast _b[M^b]\}&=&
\tilde{{\cal H}}_a
[\pounds_{M^b}
N^a]\\
\{{\cal H}[N],\tilde{{\cal H}}_a[N^a]\}&=&{\cal H}[-\pounds_{N^a}N]\\
\{{\cal H}[N],\tilde{{\cal H}}^\ast _a[N^a]\}&=&{\cal
H}^\ast [-\pounds_{N^a}N]\\
\{{\cal H}^\ast [N],\tilde{{\cal H}}_a[N^a]\}&=&{\cal
H}^\ast [-\pounds_{N^a}N]\\
\{{\cal H}^\ast [N],\tilde{{\cal H}}^\ast _a[N^a]\}&=&{\cal H}[\pounds_{N^a}N]
\label{3.15b} \eea

The remaining Poisson brackets are those that only contains $\ch $ and $\ch
^\ast $. A straightforward calculation gives
\bea \{\ch [N],\ch [M]\}&=&\ch _b[E^{bI}E^a_I(M\partial _aN-N\partial _aM)]-
\nn \\
&& \ch _b^\ast[E^{\ast bI}E^a_I(M\partial _aN-N\partial _aM)]\\
\{\ch [N],\ch ^\ast [M]\}&=&\ch _b[E^{bI}E^{\ast a}_I(M\partial _aN-N
\partial _aM)]+\nn \\
&& \ch _b^\ast[E^{bI}E^a_I(M\partial _aN-N\partial _aM)]\\
\{\ch ^\ast [N],\ch ^\ast [M]\}&=&-\ch _b[E^{bI}E^a_I(M\partial _aN-
N\partial _aM)]+\nn \\
&& \ch _b^\ast[E^{\ast bI}E^a_I(M\partial _aN-N\partial _aM)]
\eea
where I had to use the structure constant identity (\ref{id}).
Thus, the constraint algebra closes, and the total Hamiltonian
\be H^{tot}:=\int _{\Sigma}d^3x\left (N\ch +N^\ast \ch ^\ast + N^a\ch _a +
N^{\ast a}\ch ^\ast _a -A_{0I}\cg ^I \right ) \label{Htot}\ee
defines a consistent theory. Consistent, in the sense that a field
configuration that starts out on the constraint surface stays there under
time evolution. (The $\ast$'s on the Lagrange multiplier fields are just for
notational convenience.) \\ \\
Given this theory, there are of course several question to ask: {\it e.g.}
  1. What is
the interpretation of the theory, what does it describe? 2. Is this
theory diffeomorphism invariant (in the four-dimensional sense). 3. Is there a
geometrical interpretation of the fields? And, in that case, what kind of
geometry can the theory describe? Lorentzian, Euclidean, ...? 4. What is the
physical meaning of the local symmetry generated by $\ch ^\ast _a$ and $\ch
^\ast $; What are the finite symmetry transformations? The answers to all
these questions will be found from the result of the split of the theory in
the next subsection.

\subsection{$so(1,3;C) = so(3;C) \oplus so(3;C)$}\label{split}
In this subsection, I will use the fact that the Lorentz Lie-algebra splits
into two copies of complex $so(3)$ algebras, to decompose the $so(1,3)$
theory. This splitting is performed by introducing self dual and antiself
dual fields, and
since the square of the duality-operation, in Minkowskian spacetime, equals
-1, self duality necessarily means introducing complex numbers. If we instead
would use the Lie-algebras $so(4)$ or $so(2,2)$, we could still get by with
real fields, and the final result would be completely different. \\ \\
In the vector notation, introduced in the previous subsection, self duality
and antiself duality mean;
\be q^{\ast J}_IV^{(\pm)}_J=\pm i V^{(\pm)}_I \ee
for a self dual field $V^{(+)}_I$ and antiself dual field $V^{(-)}_I$. Given
any field $V_I$ that take values in the Lie-algebra $so(1,3)$ -- in the vector
representation -- the (anti)self dual part is defined as follows
\be V^{(\pm)}_I:=\d ^{(\pm)J}_IV_J=\frac{1}{2}(\d ^J_I\mp i g^{\ast J}_I)V_J
\ee
{}From this it is obvious that $V_I=V^{(+)}_I+V^{(-)}_I$, and that the dual of
$V_I$ is $V^\ast _I:=g^{\ast J}_IV_J=i(V^{(+)}_I-V^{(-)}_I)$. The (anti)self
dual projection operator has the following features:
\be \d ^{(+)J}_I\d ^{(+)K}_J=\d ^{(+)K}_I \hspace{10mm} \d ^{(+)J}_I
\d ^{(-)K}_J=0 \hspace{10mm} \d ^{(-)J}_I\d ^{(-)K}_J=\d ^{(-)K}_I \ee
Furthermore, since these projection operators are built out of invariant forms
of the $so(1,3)$ algebra, it follows that
\bea \d ^{(+)J}_If_{JKL}&=&\d ^{(+)J}_I\d ^{(+)M}_Kf_{JML}=\d ^{(+)J}_I
\d ^{(+)M}_K\d ^{(+)N}_Lf_{JMN}:=f^{(+)}_{IKL} \\
\d ^{(-)J}_If_{JKL}&=&\d ^{(-)J}_I\d ^{(-)M}_Kf_{JML}=\d ^{(-)J}_I
\d ^{(-)M}_K\d ^{(-)N}_Lf_{JMN}:=f^{(-)}_{IKL} \\
\d ^{(+)J}_I\d ^{(-)M}_Kf_{JML}&=&0 \\
f_{IJK}&=&f^{(+)}_{IJK}+f^{(-)}_{IJK} \eea
This is all that is needed to split the theory into two parts. First I split
the
fundamental fields
\be E^{aI}=E^{(+)aI}+E^{(-)aI}\hspace{15mm}A_{aI}=A^{(+)}_{aI}+A^{(-)}_{aI}
\ee
Putting this into the Hamiltonian (\ref{Htot}), yields
\be H^{tot}=\int _{\Sigma}d^3x \left (N^{(+)}\ch ^{(+)} + N^{(+)a}\ch
^{(+)}_a - A_{0I}^{(+)}\cg ^{(+)I} + c.c\right ) \label{Htot+} \ee
where
\bea N^{(+)}&:=&N+i N^\ast \\
N^{(+)a}&:=&N^a+iN^{\ast a} \\
A_{0I} ^{(+)}&:=&\d ^{(+)J}_IA_{0J} \\
\ch ^{(+)}&:=& \frac{1}{4}\e
_{abc}f^{(+)}_{IJK}E^{(+)aI}E^{(+)bJ}(B^{(+)cK}+\frac{2 \L i}{3}E^{(+)cK})
\approx 0 \\
\ch ^{(+)}_a&:=&\frac{1}{2}\e _{abc}E^{(+)bI}B^{(+)c}_I \approx 0 \\
\cg ^{(+)}_I&:=&\cd _aE^{(+)aI}=\partial _aE^{(+)aI} +
f^{(+)I}{}_{JK}A^{(+)J}_a E^{(+)aK} \approx 0 \eea
We also have
\bea &&\{A^{(+)}_{aI}(x),E^{(+)bJ}(y)\}=\d _a ^b\d ^{(+)J}_I \d
^3(x-y), \hspace{5mm} \{A^{(-)}_{aI}(x),E^{(-)bJ}(y)\}=\d _a ^b\d ^{(-)J}_I \d
^3(x-y), \nn \\
&&\{A^{(-)}_{aI}(x),E^{(+)bJ}(y)\}=\{A^{(+)}_{aI}(x),E^{(-)bJ}(y)\}=0 \eea
And since all the fields in the $so(1,3)$ theory were taken real, the complex
conjugate of the self dual part equals the antiself dual part. Thus, we see
that the $so(1,3)$ theory really equals the complex Ashtekar formulation {\it
without} the reality conditions. That is; complex Einstein gravity.
If we had started with the algebras $so(4)$
or $so(2,2)$ instead of $so(1,3)$, we would have found two identical copies
of the same real $so(3)$ or $so(1,2)$ theory. That is two copies of Einstein
gravity for Euclidean or "ultra-hyperbolic" spacetime. \\ \\
With this knowledge, that the $so(1,3)$ theory just is a real
formulation of complex Einstein gravity, it is an easy task to answer all the
questions that were asked in the end of the previous subsection:
1. The theory describes complex Einstein gravity. 2. The theory is
four-dimensionally diffeomorphism invariant. 3. The geometrical
interpretation of the fields is that
\be \tilde{g}^{\a \b}=\sqrt{-g}g^{\a \b} = \left (\begin{array}{cc}
-\frac{1}{N^{(+)}} & \frac{N^{(+)a}}{N^{(+)}} \\ \frac{N^{(+)a}}{N^{(+)}} &
-\frac{N^{(+)}}{8}E^{(+)aI}E^{(+)b}_I-\frac{N^{(+)a}N^{(+)b}}{N^{(+)}}
\end{array}
\right ) \ee
is the densitized complex spacetime metric. This follows from the
equivalence with the Ashtekar formulation, and formula for the spacetime
metric in Ashtekar's variables. See \cite{thesis}.
4. The symmetries generated by
$\ch ^\ast _a$ and $\ch ^\ast$ is complex diffeomorphisms. Therefore, any two
metrics that only differ by a complex coordinate transformation are to be
considered physically equivalent. To see this, consider {\it e.g.} the
transformations generated by $\tilde{\ch}_a^\ast$ on $E^{(+)aI}$.\\ \\

How is then the Ashtekar formulation of conventional real Einstein gravity
embedded into this larger $so(1,3)$ theory? We know that the reality
conditions that are used to select real general relativity from Ashtekar's
variables are \cite{Ash1}:
\bea Im(E^{(+)aI}E^{(+)b}_I)=0,&& \hspace{10mm}
Im\left (E^{(+)cM}f^{(+)}_{MIN}E^{(+)
(bI}\cd _cE^{(+)a)N}\right ) =0\\ Im(N^{(+)})=0, &&\hspace{10mm}
Im(N^{(+)a})=0 \eea
These conditions are easily translated back into the real $so(1,3)$
formulation where they become
\be E^{aI}E^{\ast b}_I=0, \hspace{10mm} E^{\ast c M}f_{MIN}E^{(bI}\cd _c
E^{a)N}=0, \hspace{10mm} N^\ast =0, \hspace{10mm} N^{\ast a}=0
\label{real2}\ee
 This means that any field configuration that satisfies (\ref{real2})
initially will continue to do so under the time evolution.
Note, however, that there is nothing automatic about the Lorentzian signature
of the metric, neither in the Ashtekar formulation nor in this $so(1,3)$
theory. The Lorentzian case is selected by requiring
$Re(E^{(+)aI}E^{(+)b}_I)<0$ or $E^{aI}E^b_I<0$ (By this, I mean all
eigenvalues are negative definite.)

Note also the similarity between the "reality conditions" (\ref{real2}) and
the second class constraints found in the Hamiltonian formulation of the
Hilbert-Palatini Lagrangian \cite{Ash1}, \cite{thesis}. The first two
conditions in (\ref{real2}) equal exactly these second class
constraints, while the two last conditions in (\ref{real2}) imply that $\ch
^\ast$ and $\ch _a^\ast$ drop out from the total Hamiltonian.
\subsection{The Lagrangians}\label{Lso(1,3)}
Here, I will derive two Lagrangians whose Hamiltonian formulations
equal the one given by (\ref{Htot}). This is actually a rather trivial task
once one knows that the $so(1,3)$ theory splits into the complex Ashtekar
formulation plus its complex conjugate: the total Lagrangians will just be the
complex Lagrangians that give the complex Ashtekar formulation,
plus their complex conjugate. The first order Lagrangian is the
selfdual Hilbert-Palatini Lagrangian \cite{Samuel}, \cite{JS},
\cite{thesis}, and the second order Lagrangian is the CDJ-Lagrangian
\cite{CDJ}, \cite{PP}, \cite{thesis}.
However, I will also show what the manifestly
real $so(1,3)$ Lagrangians look like, and give the definition of the
spacetime
metric in terms of the configuration space variables. \\ \\
The first order Lagrangian leading to the complex Ashtekar formulation
is
\be \cl =\frac{1}{2}\e ^{\a \b \g \d}e_{\a A}e_{\b B}R_{\g
\d}^{(+)AB}({\scriptstyle A_\e ^{CD}})
\label{sdhp} \ee
where $e_{\a A}$ is the complex tetrad and $R_{\g \d}^{(+)AB}$ is the self
dual
part of the Riemann tensor. I will assume that the full spin-connection $A_\a
^{AB}$ is real, meaning that the self dual and antiself dual connection will
be related via complex conjugation. It may seem strange to allow the tetrad
to be complex while the $so(1,3)$ connection remains real. The reason why
this is possible is that the important connection in (\ref{sdhp}) really is
the self dual $so(1,3)$ connection which necessarily is complex. Furthermore,
as is easily proven, a generic self dual field can always be written as the
self dual part of a {\it real} field, without loss of generality.

To find the manifestly real formulation, I need to add the complex conjugate
of the Lagrangian to the Lagrangian
\be \cl ^{tot}=\cl + \bar{\cl}=\frac{1}{2}\e ^{\a \b \g \d}e_{\a A}
e_{\b B}R_{\g
\d}^{(+)AB}({\scriptstyle A_\e ^{CD}})+\frac{1}{2}\e ^{\a \b \g
\d}\bar{e}_{\a A}
\bar{e}_{\b B}R_{\g
\d}^{(-)AB}({\scriptstyle A_\e ^{CD}}) \label{Ltot} \ee
where $\bar{e}_\a ^A$ is the complex conjugate of the tetrad. Now, I define
the real objects
\bea R_{\a \b}{}^{AB}&=&R^{(+)}_{\a \b}{}^{AB}+R^{(-)}_{\a
\b}{}^{AB},\hspace{10mm} R^\ast _{\a \b}{}^{AB}=i \left (
R^{(+)}_{\a \b}{}^{AB}-R^{(-)}_{\a
\b}{}^{AB}\right ) \nn \\
V_\a ^A&:=&\frac{1}{2}\left (e_\a ^A + \bar{e}_\a ^A\right ),\hspace{15mm} W
_\a ^A:=\frac{-i}{2}\left (e_\a ^A - \bar{e}_\a ^A\right ) \label{def} \eea
With this, the Lagrangian becomes
\be \cl ^{tot}=\frac{1}{2}\e ^{\a \b \g \d}\left (V_{\a A}V_{\b B} - W_{\a
A}W_{\b B}\right )R_{\g \d}{}^{AB}({\scriptstyle A_\e ^{CD}}) +
\e ^{\a \b \g \d}V_{\a A}W_{\b B} R^\ast _{\g \d}{}^{AB}
({\scriptstyle A_\e ^{CD}}) \label{lreal} \ee
which is the manifestly real $so(1,3)$ Lagrangian for complex gravity. The
complex metric is given by
\be g_{\a \b}=e_{\a A}e_\b ^A=V_{\a A}V_\b ^A - W_{\a A}W_\b ^A + i V_{(\a A}
W_{\b )}^A \ee
There is one strange feature of the Lagrangian (\ref{lreal}). If one counts
the number of components in $V_{\a A}$ and $W_{\a A}$ one gets 32, while the
corresponding fields in the Hamiltonian formulation -- $N$, $N^\ast$, $N^a$,
$N^{\ast a}$ and $E^{aI}$ -- only have 26 components. There are six components
missing! A similar loss happens when one goes from the self dual
Hilbert-Palatini Lagrangian to the Ashtekar formulation. In the Lagrangian,
the tetrad $e_{\a A}$ has 16 components, but at the Hamiltonian level, we
only find -- $N$, $N^a$ and $E^{ai}$(so(3)) -- 13 components. Three are
missing! For that case, the three missing components are the ones that can be
gauged away via antiself dual Lorentz transformations. That is, the tetrad
transforms under the full Lorentz group, while the self dual antisymmetric
products of two tetrads only transforms under the self dual part of the
Lorentz Lie-algebra. Therefore, without restricting the Hamiltonian
formulation, one may remove three components of the tetrad. Does there exist
a similar local symmetry for the Lagrangian (\ref{lreal})? Yes, besides the
normal Lorentz symmetry we also have local invariance under
\bea \d V_{\a A}&=&\L^\ast _A{}^BW_{\a B},\hspace{10mm} \d W_{\a A}=-\L^\ast
_A{}^BV_{\a B}
\nn \\
\d R_{\g \d}{}^{AB}&=&\L ^A{}_CR_{\g \d}{}^{CB} - \L ^B{}_C R_{\g
\d}{}^{CA}
\eea
where $\L ^A{}_B$ belongs to $so(1,3)$, and the $\ast$ denotes the dual
of the field: $\L^\ast _{AB}:=\frac{1}{2}\e _{AB}{}^{CD}\L _{CD}$.
Thus, we have a six-dimensional
local symmetry for the Lagrangian (\ref{lreal}) which is  {\it
not} represented by a
first class constraint at the Hamiltonian level. This ends the discussion
about the first order Lagrangian.\\ \\

Finding a second order Lagrangian corresponding to a given Hamiltonian, means
performing a Legendre transform. That is solve the momentum's equation of
motion to get a solution for the momentum in terms of the phase space
coordinates and their velocities. When there are constraints present, one
also has the choice of treating the constraints as primary or secondary w.r.t
the wanted Lagrangian. If the constraints are taken to be primary, the
solution for the momenta should automatically satisfy the constraints, while
in the case of secondary constraints no such algebraic features are required.
In Legendre transforms from the Ashtekar formulation it has been
found convenient
to treat the vector constraint as primary while the Hamiltonian and the Gauss
law constraint are considered to be secondary. For more details regarding
this, see {\it e.g.} \cite{PP}. Thus, we use the (\ref{Htot+}) version of the
$so(1,3)$ Hamiltonian and calculate the equations of motion for $E^{(+)aI}$:
\bea
F^{(+)I}_{0a}&=&\frac{N^{(+)}}{2}\e _{abc}E^{(+)bJ}B^{(+)cK}f^{(+)I}{}_{JK}
+ \frac{1}{2}N^{(+)b}\e _{bac}B^{(+)cI} \label{eom}\\
\ch ^{(+)}_a&=&\frac{1}{2}\e _{abc}E^{(+)bI}B^{(+)c}_I=0 \label{vec}
 \eea
The equations for the antiself dual fields are exact copies of (\ref{eom})
and (\ref{vec}).
To solve these equations for $E^{(+)aI}$, I will assume that the magnetic
field is non-degenerate, or $det(B^{(+)aI}B^{(+)b}_I)\neq 0$. If this is
true, I can always expand any vector field in terms of $B^{(+)aI}$.
For instance
\be E^{(+)aI}=\J ^{IJ}B^{(+)a}_J. \ee
Note that $\J ^{IJ}$ is self dual in both indices implying that the maximal
rank of it is three. Using this expansion in (\ref{vec}) shows that $\J
^{IJ}$ must be symmetric. This, together with the formula
$f^{(+)}_{IJK}B^{(+)aI}B^{(+)bJ}B^{(+)cK}=\mid B^{(+)dI}\mid \e ^{abc}$, where
$\mid B^{(+)dI}\mid :=\frac{1}{6}f^{(+)}_{IJK}B^{(+)aI}B^{(+)bJ}B^{(+)cK}\e
_{abc}$, makes it possible to rewrite (\ref{eom}) as
\be \Omega ^{(+)MI}=\frac{1}{\eta ^{(+)}}\left ( \J ^{MI} -
g^{(+)MI}Tr(\J)\right ) \label{s45} \ee
where I have defined $\eta ^{(+)}:=-(2 N^{(+)} \mid B^{(+)dI}\mid )^{-1}$ and
$\Omega ^{(+)MI}:=2 B^{(+)a(M}F^{(+)I)}_{0a}=\e ^{\a \b \d \e}F^{(+)M}_{\a
\b} F^{(+)I}_{\d \e}$. Eq. (\ref{s45}) is really just six of the nine
components of eq. (\ref{eom}). The remaining three components can be used to
fix $N^{(+)a}$, but that does not make any difference in the Legendre
transform since the choice of treating $\ch _a$ and $\ch ^\ast _a$ as primary
means that $N^a$ and $N^{\ast a}$ completely drop out from the Lagrangian.
Now, (\ref{s45}) is easily solved for $\J ^{IJ}$, and the solution for
$E^{(+)aI}$ becomes
\be E^{(+)aI}=\eta ^{(+)}\left (\Omega ^{(+)IJ}-\frac{1}{2}Tr(\Omega
^{(+)})\right )B^{(+)a}_J \label{mom} \ee
The Lagrangian is
$\cl=E^{(+)aI}\dot{A}^{(+)}_{aI}+E^{(-)aI}\dot{A}^{(-)}_{aI}- H^{tot}$, and
using (\ref{mom}) for $E^{(+)aI}$ and the complex conjugate of (\ref{mom})
for $E^{(-)aI}$, I get
\be \cl=\frac{\eta ^{(+)}}{8}\left ( \Omega ^{(+)IJ}\Omega ^{(+)}_{IJ}-
\frac{1}{2}(\Omega ^{(+)I}_I)^2\right ) + c.c \label{L1} \ee
where as before, the complex conjugate of a self dual field equals the
antiself dual field.\\ \\
To find the real $so(1,3)$ Lagrangian from here, I need some formulas
relating $\Omega ^{(+)IJ}$ and $\Omega ^{(-)IJ}$ to $\Omega ^{IJ}:=\e ^{\a \b
\d \e}F_{\a \b}^IF_{\d \e}^J$ and $\Omega ^{\ast IJ}:=
\Omega ^{IK}g^{\ast J}_K$,
etc. By definition
\be \Omega ^{(+)IJ}=\Omega ^{KL}\d ^{(+)I}_K \d ^{(+)J}_L=\frac{1}{4}\left
(\Omega ^{IJ} - i {}^\ast\Omega ^{IJ} - i \Omega ^{\ast IJ} - {}^\ast \Omega
^{\ast IJ}\right ). \ee
Using this in the Lagrangian (\ref{L1}), yields
\bea \cl&=&\frac{\eta}{8}\left ( Tr(\Omega ^2)-Tr(\Omega ^\ast \Omega ^\ast)
-\frac{1}{2}(Tr\Omega )^2 + \frac{1}{2} (Tr\Omega ^\ast )^2\right ) + \nn \\
&&\frac{\eta ^\ast}{4}\left (Tr(\Omega\Omega ^\ast )-\frac{1}{2}Tr\Omega
Tr\Omega ^\ast \right ) \label{L2} \eea
where
\bea \eta &:=&\frac{1}{2}Re(\eta ^{(+)})=\frac{1}{4}(\eta ^{(+)} + \eta
^{(-)})= -\frac{1}{2}\frac{\mid B^{aI}\mid N + \mid B^{\ast aI}\mid N^\ast
}{(\mid B^{aI}\mid ^2 + \mid B^{\ast aI}\mid ^2)(N^2 + (N^\ast )^2)} \nn
\\
\eta ^\ast &:=&\frac{1}{2}Im (\eta ^{(+)})=\frac{1}{4i}(\eta ^{(+)} - \eta
^{(-)}) = -\frac{1}{2}\frac{\mid B^{\ast aI}\mid N - \mid B^{aI}\mid N^\ast
}{(\mid B^{aI}\mid ^2 + \mid B^{\ast aI}\mid ^2)(N^2 + (N^\ast )^2)}\nn \\
Tr(\Omega ^\ast \Omega ^\ast)&:=&\Omega ^{IJ}\Omega ^{KL}g^\ast _{IK}g^\ast
_{JL},\hspace{5mm} etc \eea
This completes the Legendre transform. The Lagrangian density
(\ref{L2}) is the Lagrangian whose Hamiltonian is given by (\ref{Htot}).
Furthermore, from the explicit form of the Lagrangian, it is obvious that the
theory is diffeomorphism invariant.\\ \\
When it comes to identifying a spacetime metric in this theory, we can again
rely on the results from the conventional Ashtekar/CDJ-formulation. See for
instance \cite{CDJ}, \cite{PP}. The spacetime metric is
\be \tilde{g}^{\a \b}=\sqrt{-g}g^{\a \b}=\frac{2\eta ^{(+)}}{3} K^{(+)\a \b}
\ee
where
\be K^{(+)\a \b}:= \e ^{\a \g \d \e}\e ^{\b \k \t \r}F_{\g \d}^IF_{\e \r }
^JF_{\k \t}^Kf^{(+)}_{IJK} \ee

\section{The loop-algebra theory}\label{loop}

In the search for a Lie-algebra that provides us with the structure constant
identity (\ref{genid}), one is naturally led to extensions of the finite
dimensional Lie-algebras ($so(3)$, $so(1,3)$, etc.) that are known to satisfy
the identity. The loop-algebras are infinite dimensional
examples of such extensions.

In this section, I will study an Ashtekar-like canonical theory based on the
loop-algebra $\widetilde{so(3)}$. The construction can easily be generalized
to the algebras $so(1,2)$, $so(4)$, $so(1,3)$, as well as isomorphic ones. In
subsection \ref{alg}, I give a short description of the algebra. Subsection
\ref{Hso(3)} gives the Hamiltonian description, and, finally, in subsection
\ref{s1} I show that the theory can be seen as an infinite number of copies
of the Ashtekar formulation.

\subsection{The loop-algebra}\label{alg}

For details regarding loop-algebras and Kac-Moody algebras, see {\it
e.g} \cite{KacMoody}.\\ \\
Given a finite dimensional Lie-algebra $\ca$, there exist an infinite
dimensional extension of it $\tilde{\ca}$. If $T_I$ are the generators
of the finite dimensional base-algebra, they satisfy the relation
$[T_I,T_J]=f_{IJ}{}^KT_K$. In
the extended algebra, we have the generators $T_{I}^n$, where
$T_I^0=T_I$ and the new index $n$ denote the level of the generator, it
takes all integer values. In the extended algebra, the Lie-bracket is
\be [T_I^n ,T_J^m ]=f^{n m}{}_k f_{IJ}{}^KT_K^k =f_{IJ}{}^KT_K^
{n
+ m} \label{Lb}\ee
where we see that the structure constant $f^{n m}{}_k $ is
non-vanishing only for $n + m =k$. One may notice that it is not
possible to stop at any finite level, except the zeroth level, since
the Lie-bracket produces higher levels. However, one is allowed to use
only the even levels since the sum of two even integers is even.
Often, in the context of loop-algebras, one consider the central extension
of the algebra. In that case,
the right-hand side of (\ref{Lb}) gets an additional term $n
\d ^{n, -m}g_{IJ}c$, where $c$ is the one dimensional center -- it commutes
with all elements -- and
$g_{IJ}$ is a bilinear form of the base Lie-algebra. Furthermore, it is
straightforward to check that the centrally extended loop-algebra also allows
an extension with an element $d$:
\be [d, T_I^n]=nT_I^n,\hspace{10mm} [d,d]=0,\hspace{10mm} [c,d]=0 \ee
The resulting Lie-algebra $\widehat{\ca}=\tilde{\ca}\oplus \car c \oplus \car
d$ is called an affine Kac-Moody algebra.

Now, the name loop-algebra comes from the fact that the algebra admits a
very nice
interpretation;
it can be seen as $\ca$-algebra valued fields on the
circle, $S^1$. That is, given a $\ca$-algebra valued field $h(\q )$ that is a
function of a coordinate $\q$ on $S^1$ we may
make a fourier series expansion of it ($h(\q + 2 \p )=h(\q
)$)\footnote{Sometimes one considers a slightly weaker periodicity condition
$h(\q + 2\p )=\tau(h(\q ))$, where $\tau$ is an (outer) automorphism of the
Lie-algebra $\ca$. The resulting algebra is called a twisted
loop-algebra/Kac-Moody algebra}:
\be h(\q )=\sum_{n=-\infty}^\infty h_n \exp (-i n \q), \hspace{10mm}
h_n:=\frac{1}{2\p}\int_0^{2\p}d\q h(\q )\exp (i n \q) \ee
Thus, we see that we may write $h(\q )=h^I_nT_I^n:=h^I_nT_I \exp (-i n \q)$,
and from there, the commutation relation (\ref{Lb}) follows immediately.
We also see that $d$ can be represented as $i\frac{d}{d\q}$. This
way of representing $\tilde{\ca}$ will be used in subsection \ref{s1} to get
a better understanding of the theory. Note, however, that we are not forced
to use this representation, we may just see the Lie-algebra $\tilde{\ca}$ as
abstractly defined by (\ref{Lb}).

Now, we
need an invariant bilinear form. Here, I will only study the loop-algebra
$\tilde{\ca}$, and at the end comment on the case were one extends it with
$c$ and $d$. As in the previous section, the form is
invariant if
\be <[C,A],B>+<A,[C,B]>=0 \label{inv}, \ee
where $<\cdot ,\cdot>$ and $[\cdot ,\cdot ]$ denote the bilinear form and
the Lie-bracket, respectively. $A$, $B$, and $C$ belongs to $\tilde{\ca}$
Translating this condition to the structure constants, it becomes:
\be f^{n m}{}_k f_{IJ}{}^K\widetilde{g}^{k l}_{KL}+ f^{n l}{}_k
f_{IL}{}^K
\widetilde{g}^{k m}_{KJ}=0 \label{inv2}\ee
where I have defined $\widetilde{g}^{n m}_{IJ}:=<T_I^n ,T_J^m >$.
Using that $f^{n m}{}_k=\d (n + m -k)$ eq. (\ref{inv2}) gives that
the bilinear form must be of the form $\widetilde{g}^{n
m}_{IJ}=\tilde{g}^{n m}g_{IJ}$ where $g_{IJ}$ is an invariant bilinear
form for $\ca$, the invariance condition becomes:
\be f^{n m k} - f^{n k m} =0 \label{k2}\ee
Here, I have raised the "level-index" with $\tilde{g}^{n m}$. Since we
know that $f^{n m}{}_k $ has the value one if $n + m =k$, and is
zero otherwise, we can easily transfer the condition (\ref{k2}) to
$\tilde{g}^{n m}$:
\be \tilde{g}^{n , m}=\tilde{g}^{n + k ,m - k} \label{k5} \ee
If one sees $\tilde{g}^{n m}$ as an infinite dimensional matrix, the equation
(\ref{k5}) says that all elements on the same "cross-diagonal" must be equal.
The standard choice for this bilinear form is $\tilde{g}^{n m}=\d (n
+m)$, which satisfy the required relation (\ref{k5}). Furthermore, if we
introduce the extensions $c$ and $d$, this choice of bilinear form is unique
if we require invariance under conjugation with $d$. I will, however,
not make this choice here, instead I notice that there exist a basis for all
invariant bilinear forms. That is, every invariant bilinear form
$\tilde{g}^{n m}$ may be written as
\bea \tilde{g}^{n m}&=&\sum _{k=-\infty}^\infty a_{\lk }\hg _{\lk }^{n m}
\\
\tilde{g} _{\lk }^{n m}&=&\d (-k+n +m ) \label{gk}\eea
where the $a_{\lk }$ are taken to be real numbers. Notice the similarity
between $f^{nm}{}_k$ and $\tilde{g}_{\lk}^{nm}$. Thus, every bilinear
invariant form of the Lie-algebra may be written as
\be \tilde{g}^{nm}=v^kf^{nm}{}_k \label{gf} \ee
for some Lie-algebra valued field $v^k$. To be able to raise and lower the
"level-indices" $n$, $m$, ..., I pick out one bilinear form and its inverse
to be responsible
for that operation: $\tilde{g}_{\l0}^{nm}=f^{nm}{}_0$ and $\tilde{g}_{\l0
nm}\tilde{g}_{\l0}^{mk}=\d ^k_n$ {\it E.g.}
\be f^{nmk}:=f^{nm}{}_l\tilde{g}_{\l0}^{lk},\hspace{10mm}\tilde{g}_{\lk n}^m:=
\tilde{g}_{\lk}^{km}\tilde{g}_{\l0 nk} \ee
The following relations will then hold
\be \tilde{g}_{\lk}^{nm}=\d (-k+n+m),\hspace{10mm}\tilde{g}_{\lk n}^m=\d
(k+n-m),\hspace{10mm} \tilde{g}_{\lk nm}=\d (k+n+m) \label{hold}\ee

Translating all these bilinear forms to the $S^1$-representation, we see that
the bilinear form $\tilde{g}_{\lk}^{nm}$ corresponds to integrating over the
circle with the weight-factor $\exp (i k \q)$. That is, if $h(\q )$ and
$f(\q )$ belongs to $\tilde{\ca}$, we get
\be <h(\q ),f(\q
)>_{\lk}=h^I_nf^J_m<T_I^n,T_J^m>_{\lk}=h^I_nf^J_m\tilde{g}_{\lk}^{nm}g_{IJ}=
\sum_{n=-\infty}^\infty h_{Jn}f^J_{k-n} \ee
which corresponds to
\be <h(\q ),f(\q)>_{\lk}= \frac{1}{2\p}\int_0^{2\p}d\q Tr(h(\q )f(\q ))
\exp (i k
\q ) \label{inner}\ee
where I have written the bilinear form in $\ca$ as a trace:
$<T_I,T_J>:=Tr(T_IT_J)$. From (\ref{inner}) it is also clear that the
bilinear form is not invariant under conjugation with $d$:
\bea \d ^d<h(\q ),f(\q )>_{\lk}&&=<[d,h(\q )],f(\q )>+<h(\q ),[d,f(\q )>=
i<h'(\q
),f(\q )>+ \nn \\
&& i<h(\q ),f'(\q )>=\frac{i}{2\p}\int_0^{2\p}d\q Tr\left (h'(\q
)f(\q )+h(\q )f'(\q )\right )\exp (i k \q )= \nn \\
&&k<h(\q ),f(\q )>_{\lk} \eea
Thus, we see that the bilinear form is only invariant under
$d$-transformations iff $k=0$. For $k\neq 0$, we instead get something like a
"covariant bilinear form".\\ \\
Finally, before I start constructing the theory, I need to show that this
algebra provides a structure constant identity of the type (\ref{genid}). To
do this, I first pick a base-algebra $\ca$ that by itself has structure
constants that satisfy this type of identity. The simplest example is $\ca
=so(3)$, where $f_{IJ}{}^Kf_{KLM}=\d _{IL}\d _{JM}-\d _{JL}\d _{IM}$ with a
proper choice of basis. For $\widetilde{so(3)}$, we get
\be f_{IJ}{}^Kf_{KLM}f^{nm}{}_kf^{klj}=(\d _{IL}\d _{JM}-\d _{JL}\d _{IM})
\sum_{k=-\infty}^\infty\tilde{g}_{\lk}^{nm}\tilde{g}_{\lkmi}^{lj}
\label{loopid}
\ee
which is an identity of the requested type. Furthermore, from (\ref{hold}) it
is clear that $\tilde{g}_{\lk n}^j\tilde{g}_{\lm}^{nl}=\tilde{g}_{\lkm
}^{jl}$, which is the second requirement that is needed; the product of two
bilinear forms must be an invariant bilinear form.

\subsection{The $\widetilde{so(3)}$ Hamiltonian}\label{Hso(3)}

In this subsection, I will construct a gauge and diffeomorphism invariant
theory based on the Lie-algebra $\widetilde{so(3)}$. The construction will be
very similar to the $so(1,3)$ theory presented in subsection \ref{Hso(1,3)}.
The major difference is that we now have an infinite number of invariant
bilinear
forms for $\widetilde{so(3)}$ compared to only two for $so(1,3)$. This will
imply the introduction of an infinite number of new constraints. I will only
treat the non-extended $\widetilde{so(3)}$ here, and the reason why I do not
include the central extension and the $d$-generator is that I have not been
able to find a closed constraint algebra for that case.\\ \\
The basic fields are an $\widetilde{so(3)}$ connection $A_{aI}^n$ and its
conjugate momentum, $E^{aI}_n$. They satisfy
\be \{A_{aI}^n(x),E^{bJ}_m(y)\}=\d ^b_a\d ^J_I\d ^n_m\d ^3(x-y) \ee
To write out the constraints, I first define: $E^{aI}_{\lk n}:=\tilde{g}_{\lk
n}^mE^{aI}_m$, and analogous for $A^{In}_{\lk a}$. (If one sees $E^{aI}_n$ as
an infinite dimensional vector of three-by-three matrices, $ E^{aI}_{\lk n}$
is just $E^{aI}_n$ shifted $k$ steps along the vector.)
Here are the constraints
\bea \cg ^{In}&:=&\cd _aE^{aIn}=\partial _aE^{aIn}+f^I{}_{JK}f^{nmk}A_{am}^J
E^{aK}_k\approx 0 \\
\ch _{\lk a} &=& \frac{1}{2}\e _{abc}E^{bn}_{\lk I}B^{cI}_n \approx 0
\label{k13} \\
\ch _{\lk}&=& \frac{1}{4}\e _{abc}f_I{}^{JK}f^n {}_{m
k}E^{bm}_{\lk J}E^{ck}_K\left (B^{aI}_n + \frac{2\L}{3}E^{aI}_n \right )
\approx 0 \label{k15} \\
 \eea
where $B^{aI}_n := \e ^{abc}F_{bcn}^I=\e ^{abc}\left (2 \partial
_bA_{cn}^I + f_{JK}{}^If^{m k}{}_n A_{bm}^JA_{ck}^K\right )$.
Now it is straightforward to go through the poisson bracket calculations in
the same manner as in section \ref{Hso(1,3)}: {\bf I.} Check that $\cg _I^n $
and
$\tilde{\ch}_{\lk a}:=\ch _{\lk a} -A_{\lk an}^I\cg _I^n$ generate gauge
transformations
and spatial
diffeomorphisms times a $k$-shift, respectively. {\bf
II.} Calculate all poisson brackets containing these two constraints by
using the transformation properties of the constraints. {\bf III.} Calculate
the remaining poisson brackets $\{\ch _{\lk }[N],\ch _{\lm}[M]\}$.\\ \\
{\bf I.} The transformations generated by $\cg ^{In}$ and $\tilde{\ch} _
{\lk a}:=\ch_{\lk a}-A_{\lk a}^{In}\cg _{In}$ are
\bea
\d ^{\cg ^{In}}E ^{aI}_m&:=&\{E ^{aI}_m,\cg ^{J}_n[\L_{J}^n]\}=
f^I{}_{JK}f_{mn}{}^l\L^{Jn}E ^{aK}_l
 \nn \label{a6}\\
\d ^{\cg ^{In}}A _{am} ^I&:=&\{A _{am} ^I,\cg ^J_n[\L_J^n]\}=-\cd _a\L^I_m
\nn \\
\d ^{\tilde{\ch}_{\lk a}}E ^{aI}_n &:=&\{E
^{aI}_n ,\tilde{\ch}_{\lk b} [N^b]\}\!=\!N^b
\partial _b
 E
^{aI}_{\lk n} \!-\! E ^{bI}_{\lk n} \partial _b N^a \!+\! E ^{aI}_{\lk
n}
\partial _b
N^b\!=\!
\pounds _{N^b}E ^{aI}_{\lk n} \nn\label{k10}
 \\
\d ^{\tilde{\ch}_{\lk a} }A _a ^{In}&:=&\{A _a
^{In},\tilde{\ch}_{\lk b} [N^b]\}\!=\!N^b\partial _b
A _{\lk a} ^{In} \!+\!
A _{\lk b}^{In}\partial _a N^b\!=\!\pounds _{N^b}A _{\lk a} ^{In}
\label{k11} \eea
which shows that $\cg ^{In}$ generates gauge transformations and
$\tilde{\ch}_{\lk a}$ generates spatial diffeomorphisms times a $k$-shift.\\
\\
{\bf II.} With this knowledge, the calculation of all Poisson
brackets containing $\cg ^I_n$ and $\tilde{\ch}^{\lk }_a$ simplifies
significantly. We know that $\cg ^I_n$,
 $\ch _\lk$ are gauge-covariant, and that all
constraints are diffeomorphism covariant. This gives
\bea
\{{\cal G}^I_n [\Lambda_I^n],{\cal G}^{Jm} [\Gamma_J^m ]\}&=&{\cal
G}^K_k
[f_{KIJ}f^{k n m}\Lambda^I_n
\Gamma^J_m ]\label{k100} \\
\{{\cal G}^I_n [\Lambda_I^n],\tilde{{\cal H}}_{\lk a} [N^a]\}&=&{\cal
G}^I_n[-\pounds_{N^a}\Lambda^{n }_{\lk I} ]\\
\{{\cal G}^I_n [\Lambda_I^n],{\cal H}^\lk [N]\}&=&0\\
\{\tilde{{\cal H}}_{\lk a} [N^a],\tilde{{\cal H}}_{\lel b} [M^b]\}&=&
\tilde{{\cal H}}_{\lkl a}
[-\pounds_{M^b}
N^a]\\
\{{\cal H}_\lk [N],\tilde{{\cal H}}_{\lel a} [N^a]\}&=&{\cal
H}_{\lkl }[-\pounds_{N^a}N]
 \eea
{\bf III.} To calculate $\{\ch _{\lk} [N],\ch _{\lm }[M]\}$,
the identity (\ref{loopid}) comes in handy. The result is

\be
\{\ch _\lm [N],\ch _\len [M]\}=\sum _{k=-\infty}^\infty \ch _{\lkmn a }
[q_{\lkmi }^{ab}(N\partial _bM-M\partial
_bN)]
 \label{HH} \ee
where I have defined $q_{\lk}^{ab}:=-E^{aI}_{\lk n}E^{bn}_I$.
This completes the constraint analysis. All constraint are first class, and
the total Hamiltonian is
\be H^{tot}=\int _{\Sigma}d^3x\left (-A_{0I}^n \cg ^I_n + \sum
_{k=-\infty}^\infty \left ( N^\lk \ch _\lk +
N^{\lk a} \ch _{\lk a}\right ) \right ) \label{Htotso(3)}\ee
where $A_{0I}^n$, $N_{\lk}$ and $N^a_{\lk}$ are Lagrange multipliers.
Thus, we see that it is indeed possible to construct a consistent
Hamiltonian formulation based on the Lie-algebra $\widetilde{so(3)}$.
However, in doing so, one is forced to include an infinite number of
constraints.
The question is now; have we gained any local degrees of freedom by enlarging
the gauge algebra from $so(3)$ to $\widetilde{so(3)}$? This is not a
trivial question to answer in this description, since we have both an
infinite number of phase
space variables as well as an infinite number of constraints. If we naively
try to calculate the number of local degrees of freedom by truncating the
theory at some finite level $N$, we seem to get more constraints than phase
space variables, meaning that we have no local degrees of freedom.
 However, we know that that is not
true, we know that the non-trivial $so(3)$  Ashtekar formulation of Einstein
gravity
is contained in this theory. So, for some reason the naive count does not
give the complete answer. One reason could be that by truncating the theory
at a finite level, we actually break the gauge invariance of the theory,
meaning that the Gauss law constraint cannot any longer be treated as a first
class constraint. (The same will also be true for some of the other
constraints.) However, as I will show in the next subsection, by rewriting
this theory as a theory with a finite number of constraints on every point on
the circle, it is clear that this theory actually has an infinite number of
degrees of freedom per point in spacetime. So, what have we gained by
enlarging the conventional Ashtekar formulation in this way? My starting
motivation for enlarging the theory in the first place was the hope of
finding a real theory for Lorentzian spacetime. Can this be accomplished with
the theory presented here? To answer that question, one really needs to know
how the metric is defined in terms of the fields in this formulation.
Remember that the need of complex fields in the conventional $so(3)$ Ashtekar
formulation, can be seen as coming from the requirement that $E^{aI}E^b_I$
should be negative definite \cite{subenoy}.  And, as is shown in the
next subsection, this $\widetilde{so(3)}$ theory is really equivalent
to an infinite number of copies of complex Einstein gravity, meaning
that the theory actually has an infinite number of complex spacetime
metrics. Also, at this
stage it is not even clear that the conventional theory of real
Einstein gravity is
contained in the {\it real} $\widetilde{so(3)}$ formulation. It is obvious
that it
is contained in the {\it complex} $\widetilde{so(3})$ theory; just put all
higher
modes in the fields to zero, and we recover the conventional complex Ashtekar
formulation. It is, however, true that conventional real Einstein gravity is
contained also in the real $\widetilde{so(3)}$ theory, and to prove that,
I will
show how we can recover the ADM-Hamiltonian from the $\widetilde{so(3)}$
Hamiltonian.

Consider a field configuration in the $\widetilde{so(3)}$ theory that is
initially of the form:
\bea E^{aI}_n&=&E^{aI}g_n, \hspace{10mm} g_n:=\frac{1}{\p n}((-1)^n-1)
\label{fc1}\\
A_{aIn}&=&\d ^0_n \G_{aI}({\scriptscriptstyle E^{bJ}})-K_{aI}g_n \\
N_{\lk}&=&N\d ^0_k \\
N_{\lk}&=&N^a\d ^0_k \\
A_{0K}^n&=&-\L _K \d ^n_0 + L_K g^n \label{fc4}\eea
where $\G _{aI}$ is the spin-connection defined to annihilate $E^{aI}$:
\be D_aE^{bI}=\partial _aE^{bI}+\G _{ac}^bE^{cI}+f^{IJK}\G _{aJ}E^{bK} - \G
_{ac}^cE^{bI}=0 \label{tf} \ee
Here I have also introduced the $E^{ai}$ compatible Christoffel-connection
$\G _{ab}^c$. Both $\G _{ab}^c$ and $\G _{ai}$ can be uniquely solved for
from (\ref{tf}) above. See for instance \cite{thesis}.
Note that $g_n$ are the Fourier coefficients of the imaginary step function:
\be \sum_{n=-\infty}^\infty g_n\exp (-i n \q )=\left \{ \begin{array}{c}
 i, \hspace{5mm} 0<\q <\p \\ -i, \hspace{5mm} \p < \q < 2\p
\end{array} \right.\ee
Now, to check that this field configuration will not change under time
evolution, and also to rewrite the constraints, we will need the identity
\be f^{nmk}g_mg_k=-1 \d ^n_0 \label{g2id} \ee
{}From (\ref{g2id}) it follows that $g_ng^n=-1$, $f^{nmk}g_ng_mg_k=0$ and
$f^{nmk}g_mg_kf_{n}{}^{jl}g_jg_l=1$. Writing down the equations of motion for
$E^{aI}_n$ and $A_{aI}^n$ we get from the Hamiltonian (\ref{Htotso(3)}), and
using the field configurations (\ref{fc1})-(\ref{fc4}), yields
\bea \dot{E}^{aI}_n&=&\cdots =-\d
^0_nf^{IJK}E^b_JE^a_KD_bN+g_nNK_{bN}E^{b[I}E^{aN]} + g_nE^{[aI}D_bN^{b]} +
\nn \\
&& \d ^0_nf^{IJK}K_{bJ}N^{[b}E^{a]}_K - g_n\L _Kf^{KIJ}E^a_J - \d ^0_n
L_Kf^{KIJ}E^a_J \label{eom1} \\
\dot{A}_{aI}^n&=&\cdots = Nf_{IJK}E^{bJ}\left
(R_{ab}^K({\scriptscriptstyle \G})g^n
+ D_{[a}K_{b]}^K \d ^n_0 - f^{KLM}K_{aL}K_{bM}g^n\right )+ \nn \\
&& N^b\left (R_{baI}({\scriptscriptstyle \G})\d ^n_0 - D_{[b}K_{a]I}g^n -
f_{IJK}K_b^JK_a^K\d ^n_0\right ) + g^nD_aL_I - \nn \\
&& \d ^n_0f_{IJK}\G _a ^J \L ^K +
g^n f_{IJK}K_a^J \L ^K + \d ^n_0f_{IJK}K_a^JL^K \label{eom2} \eea
where $R_{ab}^I:=\partial _{[a}\G _{b]}^I + f^{IJK}\G _{aJ}\G _{bK}$.
{}From (\ref{eom1}) it is clear that the parts that seem to break the initial
field configuration (\ref{fc1}) are just gauge transformations plus a term
that vanishes due to the Gauss law constraint (\ref{s100}). However, to
simplify things further, we may fix this freedom also, by choosing:
\be L^I=E^{bI}D_bN + N^bK_b^I \label{fc6} \ee
With this choice, $E^{aI}_n$ will maintain its form under the time evolution.
When it comes to $A_{aI}^n$, we must require that its zero-mode
continue to annihilate $E^{aI}$ during the time evolution. That is
\be \{ D_aE^{bI}, H^{tot}\}=D_a\dot{E}^{bI}+\dot{\G }^b_{ac}
E^{cI} - \dot{\G }^c_{ac}E^{bI} + f^{IJK}\d ^n_0\dot{A}_{aJn}E^{b}_K =0
\label{gammadot} \ee
It is straightforward, although rather tedious, to put (\ref{eom1}) and
(\ref{eom2}) into (\ref{gammadot}) and show that the initial field
configuration is indeed maintained under the time evolution. Furthermore,
using this field configuration in the constraints, we get
\bea \cg ^{In}&=&\d ^n_0f^{IJK}K_{aJ}E^a_K  \label{s100} \\
\ch _{\lk a}&=& g_{\lk}E^{bI}R_{abI}({\scriptscriptstyle \G _{cK}}) + \d
^0_{\lk}E^{bI}D_{[a}K_{b]I}-g_{\lk}E^{b}_If^{IJK}K_{aJ}K_{bK} \approx  \nn
\\
&& \d
^0_{\lk}E^{bI}D_{[a}K_{b]I} \label{v1} \\
\ch _{\lk}&=& -\d
^0_{\lk}\frac{1}{2}f_{IJK}E^{aI}E^{bJ}(R_{ab}^K({\scriptscriptstyle \G
_{cK}})-f^{KLM}K_{aL}K_{bM})+g_{\lk}f_{IJK}E^{aI}E^{bJ}D_{a}K_{b}^K+ \nn \\
&& \approx \frac{1}{2}\d ^0_{\lk}\left ( -f_{IJK}E^{aI}E^{bJ}
R_{ab}^K({\scriptscriptstyle \G _{cK}})+E^{aI} E^{bJ}K_{a[I}K_{bJ]} \right )
\label{H1} \eea
where I have used the Bianchi identity and the Gauss law in (\ref{v1}), and
the Gauss law in (\ref{H1}). But, these constraints are just the conventional
ADM-constraints for triad gravity. This means that every solution to
Einstein's
equations is also a solution to the real $\widetilde{so(3)}$ theory. However,
 the price we have to pay here to get a real Ashtekar-like theory,
containing Einstein gravity, is rather high; we had to
introduce an infinite number of constraints as well as phase space variables,
per point. Also, if the example above shows a generic feature of the
inclusion of Lorentzian gravity into the real $\widetilde{so(3)}$ theory,
we must allow the fields to have "infinitely long tails" -- all odd higher
modes have to be non-zero -- and with a very slow convergence.
Of course, the field
configurations chosen here, to show the relation to GR, is only one example,
and it may be possible to find nicer field configurations that do the job. In
fact, as I will show in the next subsection, the above mentioned problems
with this theory can be solved by considering a
 theory based on a finite dimensional version of the Lie-algebra
$\widetilde{so(3)}$.
\subsection{The $\widetilde{so(3)}$ theory on a circle}\label{s1}
As was shown in subsection \ref{alg}, one may see a general element in the
loop-algebra $\widetilde{so(3)}$ as an $so(3)$ valued function on the circle.
That is
\bea E^{aI}(\q )&:=&\sum _{n=-\infty}^\infty E^{aI}_n \exp (-i n \q ),\hspace
{5mm}\Rightarrow \hspace{5mm} E^{aI}_n=\frac{1}{2\p }\int ^{2\p }_0 d\q
E^{aI}(\q )\exp (i n \q )  \\
A_{a}^I(\q )&:=&\sum _{n=-\infty}^\infty A^I_{an} \exp (-i n \q ),\hspace
{5mm}\Rightarrow \hspace{5mm} A^I_{an}=\frac{1}{2\p }\int ^{2\p }_0 d\q
A^I_{a}(\q )\exp (i n \q )  \\
N(\q )&:=& \sum _{k=-\infty}^\infty N_{\lk} \exp (-i k \q ),\hspace
{5mm}\Rightarrow \hspace{5mm} N_{\lk}=\frac{1}{2\p }\int ^{2\p }_0 d\q
N(\q )\exp (i k \q )  \\
N^a(\q )&:=& \sum _{k=-\infty}^\infty N^a_{\lk} \exp (-i k \q ),\hspace
{5mm}\Rightarrow \hspace{5mm} N^a_{\lk}=\frac{1}{2\p }\int ^{2\p }_0 d\q
N^a(\q )\exp (i k \q )  \\
A^I_{0}(\q )&:=&\sum _{n=-\infty}^\infty A^I_{0n} \exp (-i n \q ),\hspace
{5mm}\Rightarrow \hspace{5mm} A^I_{0n}=\frac{1}{2\p }\int ^{2\p }_0 d\q
A^I_{0}(\q )\exp (i n \q ) \eea
Using this in the fundamental Poisson bracket, we get
\be \{A_{aI}(x,\q ),E^{bJ}(y,\q ')\}=\d ^b_a\d ^J_I\d
^3(x-y)\sum_{n=-\infty}^\infty \exp (-i n(\q - \q '))=2\p \d ^b_a\d ^J_I\d
^3(x-y)\d (\q -\q ') \ee
and the total Hamiltonian becomes
\be H^{tot}=\frac{1}{2\p}\int _{\Sigma}d^3x\int _0^{2\p}d\q \left ( N(\q )\ch
(\q ) + N^a(\q
)\ch _a(\q ) -A_{0I}(\q )\cg ^I(\q )\right ) \label{Htots1} \ee
where
\bea \ch (\q )&=&\frac{1}{4}\e _{abc}f_{IJK}E^{aI}(\q )E^{bJ}(\q )B^{cK}(\q )
\\
\ch _a(\q )&=&\frac{1}{2}\e _{abc}E^{bI}(\q )B^c_I(\q ) \\
\cg ^I(\q )&=&\cd _aE^{aI}(\q )=\partial _aE^{aI}(\q )+f^{IJK}A_{aJ}(\q
)E^a_K (\q ) \eea
That is we have an infinite number of copies of the conventional complex
$so(3)$ Ashtekar formulation, without any reality conditions. An infinite
number of copies of complex gravity.
 The copies are labeled by the continuous
parameter $\q $. Thus, from here it is clear that we really have an infinite
number of degrees of freedom per point in spacetime; or we have four degrees
of freedom per point in $\cm \times S^1$. This means that if we choose to see
the space $S^1$ as being a compactified fifth dimension, we really just have
four degrees of freedom per point in the five-dimensional spacetime. On the
other hand, if we insist on regarding the space $S^1$ as an abstract internal
space, just introduced for notational convenience, we get an infinite number
of degrees of freedom per point in spacetime. \\ \\
If we want to select real general relativity, we know from the
conventional Ashtekar formulation that we should have
\bea &&Im(E^{aI}(\q )E^b_I(\q ))=0,\hspace{10mm}
 Im(E^{cI}(\q )E^{(aJ}(\q )\cd _cE^{b)K}(\q )f_{IJK})=0 \nn \\
&&Im(N(\q ))=0,\hspace{10mm} Im(N^a(\q ))=0 \eea
which may be translated back to the real $\widetilde{so(3)}$ formulation as
\bea &&f^{nm}{}_kE^{aI}_nE^b_{Im}=\d ^0_k E^{aI}_nE^{bn}_I,\hspace{10mm}
 N_{\lk}=N_{\lkmi},\hspace{10mm} N^a_{\lk}=
N^a_{\lkmi} \nn \\
&& f^{nm}{}_kf^{kl}{}_jE^{cI}_mE^{(aJ}_l\cd _cE^{b)Kj}f_{IJK}=\d ^n_0
f^{kl}{}_jE^{cI}_kE^{(aJ}_l\cd _cE^{b)Kj}f_{IJK} \label{real3}  \eea
The field configuration (\ref{fc1})-(\ref{fc4}) satisfies this.\\ \\

If one is interested in knowing what the field configuration
(\ref{fc1})-(\ref{fc4}) corresponds to here in the $S^1$-formulation, it is
straightforward to Fourier expand it:
\bea E^{aI}(\q )&=&iE^{aI}f(\q ),\hspace{10mm} A_{aI}(\q )=\G
_{aI}({\scriptscriptstyle E^{bJ}})-i K_{aI}f(\q ) \\
N(\q )&=&N,\hspace{10mm}N^a(\q )=N^a \eea
where
\be f(\q )=\left \{\begin{array}{l}1,\hspace{5mm}0<\q <\p \\ -1,\hspace{5mm}
\p <\q <2\p \end{array}\right. \ee
Before leaving this theory, I want to comment on how to construct a related
theory with a finite number of degrees of freedom per point. As we have seen
above, the loop algebra $\widetilde{so(3)}$ is isomorphic to a direct sum of
an infinite number of $so(3)$ algebras. What if we just consider a direct sum
of $N$ $so(3)$ algebras? Or put differently, the loop algebra can be seen as
a map from $S^1$ to the base algebra. Alternatively we could consider a
periodic one-dimensional lattice with $N$ lattice points, and construct an
"lattice-algebra" as a map from the lattice to the base-algebra. Explicitly
one could start from $N$ $so(3)$ valued fields $A_{aI}(n)$, $E^{aI}(n)$,
$N(n)$, $N^a(n)$ and $A_{0I}(n)$ labeled by the integer $n$, and make the
lattice periodic by requiring $E^{aI}(n+N)=E^{aI}(n)$ and similarly for the
other fields. The next step to reach the "lattice-algebra" is to Fourier
expand the fields:
\be E^{aI}(n)=\sum_{k=1}^NE^{aI}_k\exp (-i k \frac{2\p n}{N}) \hspace{3mm}
\Rightarrow \hspace{3mm} E^{aI}_k=\frac{1}{N}\sum_{n=1}^NE^{aI}(n)\exp (i k
\frac{2\p n}{N}) \label{Fe} \ee
and similarly for all the other fields. In the "lattice-algebra" formulation
everything will look exactly as in the loop-algebra formulation in subsection
\ref{Hso(3)} except that we now should identify two fields that differ
by $N$ steps in the level-index: $E^{aI}_{k+N}=E^{aI}_k$ etc. Then, we could
again try to find the theory of Einstein gravity embedded in the {\it real}
"lattice-algebra" theory, and it is easily shown that for $N$ odd and $>1$,
the
following field configuration will work:
\bea  E^{aI}(n)&=&iE^{aI}f(n),\hspace{10mm} A_{aI}(n )=\G
_{aI}({\scriptscriptstyle E^{bJ}})-i K_{aI}f(n ) \\
N(n )&=&N,\hspace{10mm}N^a(n)=N^a \eea
where
\be f(n )=\left \{\begin{array}{l}1,\hspace{5mm}1\leq n \leq \frac{N}{2}-1
 \\ -1,\hspace{5mm} \frac{N}{2}+1 \leq n \leq N \end{array}\right. \ee
and the Fourier transformed field $E^{aI}_k$ etc. are given by (\ref{Fe}).

\section{Conclusions}
The fact that the $so(1,3;R)$ and the $\widetilde{so(3;R)}$ theory are shown
to be equivalent to the $so(3;C)$ Ashtekar formulation, and infinitely many
copies of it, should not come as a total surprise. That is
just a reflection of the isomorphisms
\be so(1,3;R)\cong so(3;C)\hspace{10mm} and\hspace{10mm}
\widetilde{so(3;R)}\cong \sum _{n=-\infty}^\infty \oplus so_n(3;C). \ee
The interesting
thing about this formulations is that the reality conditions are transformed
into normal constraint-like conditions on the fields. Thus, if it is possible
to keep the reality conditions "outside" the theory, and only impose them
in the quantum inner product\footnote{This is, so
far, the standard way of postponing the problem of dealing with
the reality conditions in
the quantization program of the Ashtekar variables. One assumes that the
reality conditions
may be used to select the correct inner product.}, then the same is possibly
 true for certain types of (second class) constraints. If that is the case,
what is the criteria
on the constraints, making this feasible? Guided by the real formulations in
these paper, one may make the following conjecture: if the real theory has a
complex structure, and the constraints are equivalent to requiring some
objects to be "real" w.r.t this complex structure, then it ought to be
possible to use these constraints to select an inner product.

More
specifically: Take a real vector space $\cv$ equipped with the complex
structure $J$, such that $J^2=-{\bf 1}$, and a "complex conjugation"
operator $C$, such that $C^2={\bf 1}$ and $CJC=-J$. "Complex conjugation"
is defined as $\bar{A}:=CA$ for all vectors $A$ in $\cv$. An object --
vector or function of vectors -- is said to be "real" if it is invariant
under $C$. An "imaginary" object changes sign under $C$. If we now, in our
theory, have a set of constraints that is equivalent to requiring some
objects to be "real", then we may put these constraints outside the theory,
and later use them to select an inner product. The criteria we should have
on the inner product is that the operators that correspond to the "real"
objects, should be self adjoint.

In the real theories studied in this paper, the above objects may be
specified as follows. For the $so(1,3;R)$ theory: the vector space $\cv$
is the
space of $so(1,3;R)$ valued functions on spacetime. The "real" subspace $\cv
_R$ is given by pure rotations, while $\cv _I$ contains only pure boosts.
"Complex conjugation" thus corresponds to changing sign of the boost-part of
a general element in $\cv$, and, using the basis (\ref{basis}), we may
represent $C$ as: $C=diag(-1,-1,-1,1,1,1)$.
The complex structure $J$ is given by
$g^\ast_I{}^J$. It satisfies $g^\ast_I{}^Jg^\ast_J{}^K=-\d _I^K$ and it maps
rotations into boosts and vice versa. A "real" object is invariant under
"complex conjugation", and it is easy to check that {\it e.g.} $E^{aI}E^b_I$
is "real" while $E^{\ast aI}E^b_I$ is "imaginary".

In the $\widetilde{so(3;R)}$ theory, the vector space $\cv $ is the space of
$\widetilde{so(3;R)}$ valued functions on spacetime. The "real" subspace
$\cv _R$ consists of elements that is even under $n\rightarrow -n$, while
the imaginary subspace $\cv _I$ contains the odd elements. Here, $n$ is the
infinite dimensional level-index in $\widetilde{so(3;R)}$. The complex
structure $J$ is given by $J_n{}^m:=f_n{}^{mk}g_k$, where $g_k$ is the
fourier coefficients of the imaginary step-function (\ref{fc1}). This $J$
satisfies $J_n{}^mJ_m{}^k=-\d _n^k$ and it maps even elements into odd ones,
and vice versa. "Complex conjugation" is given by the simple operation $n
\rightarrow -n$, and may be represented as: $C_n{}^m=\d (n+m)$. Thus, we
see that {\it e.g.} $E^{aI}_nE^{bn}_I$ is real, while
$E^{aI}_{\lk n}E^{bn}_I$ is
"complex".

Note that, for both these two examples, "complex conjugation" is really an
automorphism of the real Lie-algebras $so(1,3;R)$ and $\widetilde{so(3;R)}$.

This shows that the constraints (\ref{real2}) and (\ref{real3}) can be
interpreted as "reality conditions", and may thus possibly
be used to select an inner
product.
\\ \\
Note, though, that it is not yet clear that the reality
conditions, in Ashtekar variables, can be successfully used to select an
inner product. No one has yet shown how this should be done, in practice.
I just say that {\it if} that is the case, then one should also be able to
handle certain types of second class constraints the same way. (Actually,
there is no reason to just restrict this to constraints of second class; even
first class constraints might be handled this way.)\\ \\
Finally, I want to comment on my starting motivation for studying these
theories; I really looked for a generalization of Ashtekar's variables that
could serve as a unified theory of gravity and Yang-Mills theory. The reason
why the theories studied here only give complex gravity, may be understood
from the fact that both $so(1,3;R)$ as well as $\widetilde{so(3)}$ are
isomorphic to direct sums of $so(3;C)$ algebras. Therefore, to find something
more non-trivial, one needs to find a Lie-algebra whose structure constants
satisfy an identity of the requested type (\ref{genid}), but which is not
isomorphic to sums of $so(3;C)$ algebras. I do not know of any such algebras.
Also, although the theories described in this paper were shown to be
equivalent to complex gravity, it is not clear what kind of linearized
dynamics that would follow from {\it e.g.} a linearization of the real
$so(1,3)$ theory around de Sitter spacetime. It would indeed be interesting
if for instance the $su(2)$ Yang-Mills theory falls out from the linearized
theory. (Since the number of degrees of freedom in the $so(1,3)$ theory is
only four per spacetime point, the de Sitter solution has to be a very
degenerate point in the space of solutions, for this to be possible.) \\ \\

{\bf Acknowledgements} \\
I thank Abhay Ashtekar, Fernando Barbero, Ingemar Bengtsson and Madhavan
Varadarajan for interesting discussions. This work was supported by the NFR
(Sweden) contract no. F-PD 10070-300.

\end{document}